# Performance of scintillation materials at cryogenic temperatures

V. B. Mikhailik* and H. Kraus

University of Oxford, Department of Physics, Denys Wilkinson Building, Keble Road, Oxford OX1 3RH, UK

**Abstract**

An increasing number of applications of scintillators at low temperatures, particularly in cryogenic experiments searching for rare events, has motivated the investigation of scintillation properties of materials over a wide temperature range. This paper provides an overview of the latest results on the study of luminescence, absorption and scintillation properties of materials selected for rare event searches so far. These include $CaWO_4$, $ZnWO_4$, $CdWO_4$, $MgWO_4$, $CaMoO_4$, $CdMoO_4$, $Bi_4Ge_3O_{12}$, $CaF_2$, $MgF_2$, $ZnSe$ and $AL_2O_3$-Ti. We discuss the progress achieved in research and development of these scintillators, both in material preparation and in the understanding of scintillation mechanisms, as well as the underlying physics. To understand the origin of the performance limitation of self-activated scintillators we employed a semi-empirical model of conversion of high energy radiation into light and made appropriate provision for effects of temperature and energy transfer. We conclude that the low-temperature value of the light yield of some modern scintillators, namely $CaWO_4$, $CdWO_4$ and $Bi_4Ge_3O_{12}$, is close to the theoretical limit. Finally, we discuss the advantages and limitations of different materials with emphasis on their application as cryogenic phonon-scintillation detectors (CPSD) in rare event search experiments.

* corresponding author: e-mail vmikhai@hotmail.com, telephone: 44-(0)1865-273459

**Content**

## 1. Introduction

The research on scintillation materials has a long and venerable history that saw alternating periods of a great enthusiasm and disinterest. The appearance of scintillators was prompted by a need for detection of ionising radiation that followed the discovery of X-rays and natural radioactivity at the end of the 19$^{th}$ century. CaWO$_4$ [1] and ZnS [2] were the first scintillation materials introduced for practical detection of high-energy photons and particles. It was not too long before scintillators began to play an important role in explorations and eventually underpinned several important scientific breakthroughs that shaped modern science. Perhaps the most renowned is Rutherford's study of radioactivity and the discovery of atomic electronic structure [3]. In the late 1940s the first single crystal scintillator NaI-Tl was found [4] and since then the development of new scintillation materials exhibited continued progress throughout several decades [5].

The development of scintillation materials was always motivated by the application of ionising radiation across many fields of technology (medical diagnostics, well-logging, flaw detection, security measures) as well as requirements in advanced detectors for fundamental science. During the last decade there has been an increase in interest in scintillators driven by several major developments. For example, dynamical medical imaging requires very fast data acquisition. This prompted a search for high light yield, fast scintillators, which resulted in the introduction of several new Ce-doped compounds [6]. The need for scintillation detectors with high stopping power and short response time for high-energy accelerator physics has lead to the development of lead tungstate scintillators [7]. These in turn gave a strong impetus to studies of the physical processes of conversion of high-energy radiation into light in scintillation materials (see e.g. [8, 9, 10] and reference therein).

A new field of application for scintillators emerged recently through strong interest in rare event searches [11]. Finding neutrino-less double-beta decay and detecting Weakly Interacting Massive Particles (WIMP) requires detectors capable of discriminating the weak and rare signal over the dominating background of spurious events caused by natural radioactivity and cosmic rays. The Milano group [12] were first in experimentally realising an idea expressed earlier in [13]. They demonstrated that simultaneous detection of heat and a light signal by cryogenic detectors holds great promise as it allows recognition of the type of particle interaction (neutrons, $\alpha$- / $\beta$-particles or $\gamma$-quanta). This enables efficient identification of nuclear recoils (caused by WIMP and also neutron interactions) by eliminating electron recoils due to radioactive background. Thus, a new generation of hybrid cryogenic phonon-scintillation detectors (CPSD), combining powerful background discrimination with remarkable energy threshold and resolution, has been developed and introduced into experimental practice [14, 15].

The active background rejection using a combination of phonon and scintillation responses provides a well resolved signal down to very low energies ($\leq 10$ keV) and the search for rare events is the field that benefits especially from the development of this technique. Evidence for the effectiveness of the new detection technique comes from recent results obtained in experiments searching for WIMP dark matter [16], and in studies of very long-lived isotopes [17, 18] that would have been impossible to achieve without CPSDs. It is now widely agreed that CPSDs are ideal to reach the sensitivity levels required by future experiments searching for dark matter [19] and double-beta decay [20].

Inorganic scintillators are a key component of CPSD and identification, characterisation and optimisation of scintillation materials for low-temperature application are important tasks for achieving the required sensitivity. Since the publication of a first review paper on this topic [21] there has been a swift increase in research activities and the development of scintillator materials for cryogenic rare event search experiments that resulted in a number of publications [22, 23, 24, 25, 26, 27, 28, 29, 30, 31, 32, 33, 34, 35] as well as designated workshops (CryoScint [36] and RPScint [37]).

It is the aim of this paper to discuss the latest progress achieved in research and development of scintillation materials for these applications, both in material preparation and in the understanding of the scintillation mechanisms, as well as the underlying physics. The results obtained through studying optical, luminescence and scintillation characteristics of the materials selected so far for rare event searches will be presented. Systematic studies of these properties are shown to be a highly-effective method for understanding the origins of the basic performance limits of known materials. Further advances should lead to new or improved compounds for a new generation of cryogenic scintillation detectors.

## 2. Characterisation of materials for CPSD

### 2.1 Experimental

To measure scintillation characteristics of materials over a wide range of temperatures we used the multi-photon counting (MPC) technique [38, 39]. The MPC is based on recording a sequence of photoelectron pulses produced by a PMT when detecting photons from a scintillation event. Each pulse in the sequence corresponds to an individual photon impinging on the photocathode of the PMT. The distribution of arrival times of the photons provides information on the decay characteristics of the scintillation process, while the number of photons detected per event is proportional to the light yield of the scintillator. Thus, recording a large number of scintillation events ($10^3 - 10^4$) one can obtain the characteristics of decay time and light output in a single measurement.

The measurements of scintillation properties discussed here were carried out in a $^4$He-flow cryostat using an $^{241}$Am α-source (5.5 MeV). For compatibility of the experimental results we measured samples of the same dimensions ($5 \times 5 \times 1$ mm$^3$) and maintained a fixed geometry of the experiment. Provided these conditions are fulfilled, the accuracy of the measurements is assessed to be within 10%. The error on the relative measurements of the light output for different materials is 30%, as it includes additionally the errors of the measurements of luminescence spectra and the spectral sensitivity of PMT.

The spectroscopic properties of the materials, especially absorption and luminescence spectra are also of significance for scintillation applications. Studies of these characteristics can give information on the factors affecting the quality of a scintillators. The luminescence spectra presented here were measured using high-energy VUV excitation using the synchrotron radiation source at HASYLAB (DESY); they are corrected for the spectral response of the detection system. The absorption spectra were measured using a Perkin-Elmer Lambda 9 spectrophotometer.

The experimental samples discussed in this study were obtained from different sources. $CaWO_4$, $CaMoO_4$ and $ZnWO_4$ are the off-cuts of Czochralski-grown crystals used for the production of cryogenic scintillation detectors. They where supplied by SRC Carat, Lviv, Ukraine ($CaWO_4$ and $CaMoO_4$) and Institute for Scintillation Materials, Kharkiv, Ukraine ($ZnWO_4$). The Institute for Single Crystal also provided the samples of $ZnMoO_4$, $MgWO_4$ and supplied commercial samples of $Bi_4Ge_3O_{12}$, $ZnSe$ and $Al_2O_3$-Ti. The samples of $CdMoO_4$, $CdWO_4$, $CaF_2$ and $MgF_2$ where received from Hilger Crystals (Margate, UK).

## 2.2. Luminescence spectra

CaWO$_4$, ZnWO$_4$ and CaMoO$_4$ are currently materials of first choice for cryogenic dark matter search. While CaWO$_4$ and ZnWO$_4$ have a long history of practical application as detectors of ionising radiation that resulted in a significant amount of relevant information gathered by generations of scientists [40, 41, 42], the scintillation properties of CaMoO$_4$ became of interest only recently [27, 43]. Nonetheless the spectroscopic properties of these crystals have been studied for decades [44, 45, 46, 47.] and recently we investigated them extensively [48, 49].

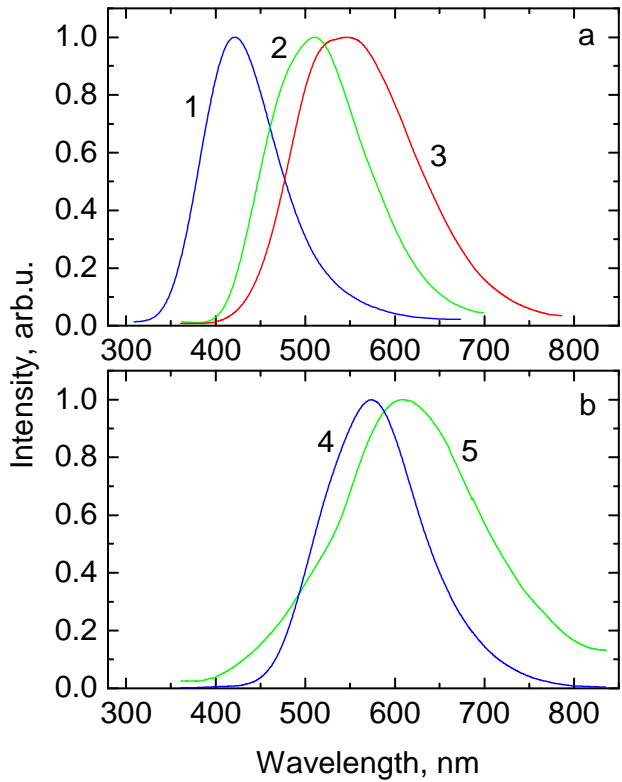

Fig. 1. Luminescence spectra of CaWO$_4$ (1), ZnWO$_4$ (2), CaMoO$_4$ (3) CdMoO$_4$ (4) and ZnMoO$_4$ (5) measured at T=8 K for excitation with 31 eV VUV photons.

High-energy excitation of the crystals under investigation gives rise to broad band emission with a maximum at 420 (CaWO$_4$) 510 (ZnWO$_4$) and 540 nm (CaMoO$_4$), as shown in Fig.1a.

Figure 1b displays the emission spectra of $CdMoO_4$ and $ZnMoO_4$ crystals that exhibit a peak at 575 and 610 nm. A common feature of the luminescence spectra of the crystals is their asymmetrical shape that indicates a composite character of the emission. Calculation studies of the electronic structure of the crystals confirmed that the upper occupied states have mostly O 2p character and the lower unoccupied band is mainly made of d-states of the W or Mo [50, 51]. Therefore, to describe the luminescence properties of tungstates and molybdates, emphasis is usually placed on the coordination number and the geometry of this complex. It is now generally accepted that the high-energy band is due to the radiative recombination of a self-trapped electron, localised at the oxyanion complex $[MeO_n]$ (Me=W or Mo) while the emission in the low-energy part of the spectrum is usually attributed to the defect, i.e. $[MeO_n]$ complex with the oxygen vacancy [52]. It is worthwhile to note that recent theoretical [53, 54] and experimental investigations [55] demonstrated that the Jahn-Teller interaction can also contribute to the complex structure of the emission spectrum in the crystals of interest.

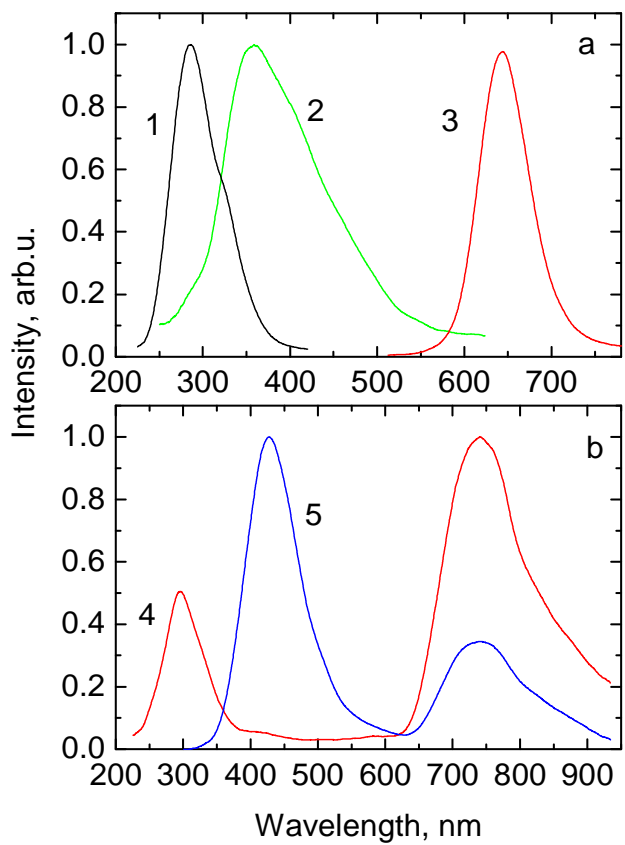

Fig. 2. (a) Luminescence spectra of $CaF_2$ (1), $MgF_2$ (2) and ZnSe (3) measured at excitation with 31 eV VUV photons. (b) Luminescence spectra of $Al_2O_3$-Ti measured at excitation with 31 (4) and 6 eV photons (5). T=8 K.

With increasing temperature the emission maxima shift towards higher energy and the luminescence spectra exhibit broadening; such thermal changes are due to the interaction of the $(MeO_n)$ emission centre with lattice vibrations [56, 49, 57]. The total intensity of the emission decreases as the temperature increases and that is explained by the effect of thermal quenching (see further discussion in chapter 2.4).

The luminescence spectra of $MgF_2$ and $CaF_2$ are displayed in figure 2a. Calcium fluoride exhibits a main emission band at 280 nm, the intensity of which strongly increases with cooling. Magnesium fluoride emits no light at room temperature; the luminescence, peaking at 385 nm, is observed only at low temperatures. The luminescence properties of binary fluorides has been studied a while ago [58, 59, 60, 61] and it is now generally accepted that the emission is due to the radiative decay of triplet self-trapped excitons.

Figure 2a shows the luminescence spectrum of undoped ZnS. The emission spectrum of ZnSe is very similar to that of ZnSe-Te, and the same concerns the luminescence kinetics of both materials [24]. According to [62] the main emission band of ZnSe-Te, peaking at 640 nm, is assigned to the radiative recombination process that occurs within the $Zn_i$-$V_{Zn}$-$Te_{Se}$ complex. In terms of this model the electron recombines with the hole captured by the hole centre ($V_{Zn}$) while Te ions substituting for Se facilitate the hole capture process at high temperatures. Since Te does not participate directly in the emission process we inferred that the luminescence properties of undoped ZnSe are due to the same mechanism, namely the radiative recombination of electrons and holes captured at zinc vacancies.

The luminescence properties of Ti-doped sapphire under high-energy excitation have been studied using different techniques [31, 63]. The luminescence spectra of the crystal (see fig. 2b) exhibit three main emission bands peaking at 290, 430 and 740 nm. The near-infrared emission band is due to the well known d-d transition in $Ti^{3+}$ ions. From the results of extensive studies of the emission and the kinetic characteristics we concluded that the ultraviolet emission band is due to the radiative decay of excitons localised at $Ti^{3+}$ ions [63]. The nature of the blue band remained a subject of extensive discussion for long, until recently investigations of the concentration dependences of the luminescence of $Al_2O_3$-Ti resulted in an important finding. It was observed that the intensity of the 430 nm band increases as the concentration of Ti decreases [31]. This provided a firm argument for the emission being an inherent property of the host matrix and it was attributed to the emission of F-centres. This emission band makes a significant contribution to the total light emitted by the scintillator at a low concentration of Ti (~10 ppm). Therefore, manufacturing crystals with a high content of anion vacancies is suggested for the improvement of the scintillation yield [31]. Focusing on increasing the light yield (number of photons emitted) in the blue emission band would favour improving the detection efficiency of sapphire CPSD. This is because the signal from a cryogenic calorimeter is proportional to the total energy it absorbs, and thus, for the same light yield, scintillation at higher photon energies is preferred..

## 2.3. Optical absorption spectra

The use of large scintillation targets implies that self-absorption of the emitted light may considerably reduce the amount of energy that reaches the light detector. This effect is especially important in crystals with a high index of refraction since more photons are trapped inside the scintillator volume and there are fewer chances for escape [64]. The optimisation of $ZnWO_4$ provides a convincing illustration of this effect: the increase of the scintillation light output by 40% was achieved by virtue of fourfold reduction of the absorption coefficient in the region of the emission band [25]. Therefore the understanding of the features of absorption in scintillating materials is vital.

As a rule, the absorption spectra of tungstate and molybdate crystals exhibit sharp absorption, characteristic of a fundamental absorption edge. It is often accompanied by a low-energy absorption tail that is attributed to defects or impurities and serves as a good indicator of material quality. The spectra displayed in figure 3 depict these features in the absorption of several crystals under study. As can be seen, two samples ($MgWO_4$ and $ZnMoO_4$) exhibit strong absorption below the fundamental absorption edge. We deduce that this is due to the presence of impurities or defects. It is worth remarking that both crystals are experimental samples that have been produced for the first time and their optimisation is only at the very beginning. Other (high quality) crystals exhibit steep absorption that can be attributed to the generic absorption edge.

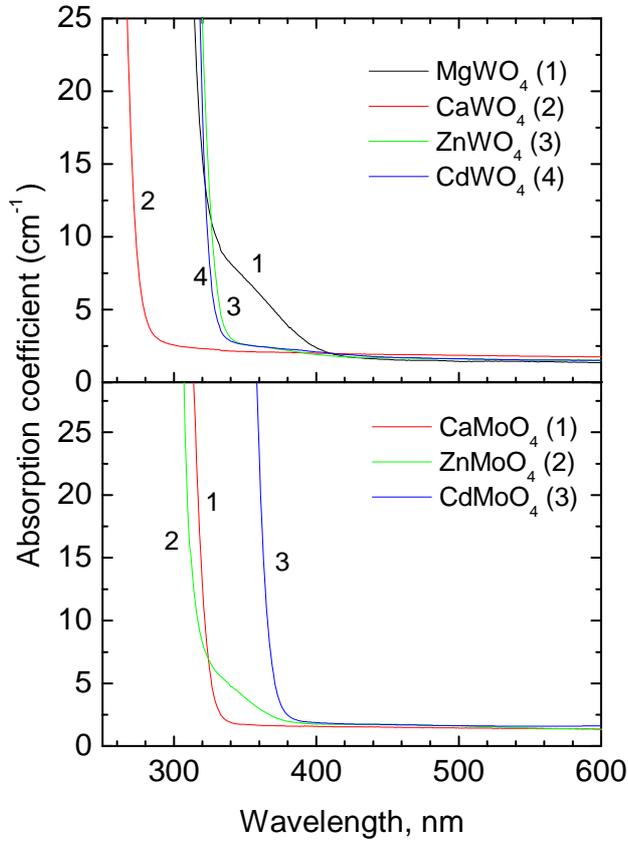

Fig. 3 (a) Absorption spectra of tungstate crystals: 1- $MgWO_4$, 2-$CaWO_4$, 3- $ZnWO_4$, 4-$CdWO_4$. (b) Absorption spectra of molybdate crystals: 1- $CaMoO_4$, 2-$ZnMoO_4$, 3- $CdMoO_4$. T=295 K.

Recently Lacomba–Peraales et al [65] discussed a correlation between the value of the band gap energy of tungstates and the ionic radius of the bivalent cation. From the data presented in fig. 3 we infer that the structure of the crystal affects the absorption more significantly than the size of the constitutive cations. Indeed, the absorption edges of the tungstates with wolframite structure ($MgWO_4$, $CdWO_4$ and $ZnWO_4$) are very close, while that of $CaWO_4$ with scheelite structure is shifted to high energies. The underlying explanation of this effect lies in the electronic structure of the bands that determine the absorption edge of the crystals. Theoretical calculations established that the top of the valence band in tungstate and molybdate crystals is dominated by O 2p states. The difference arises from the structure of the bottom of the conduction band. In CaWO$_4$ it is formed by W 5d states located at higher energies whereas in wolframite there is noticeable contribution from the s-states of bivalent cations [51].

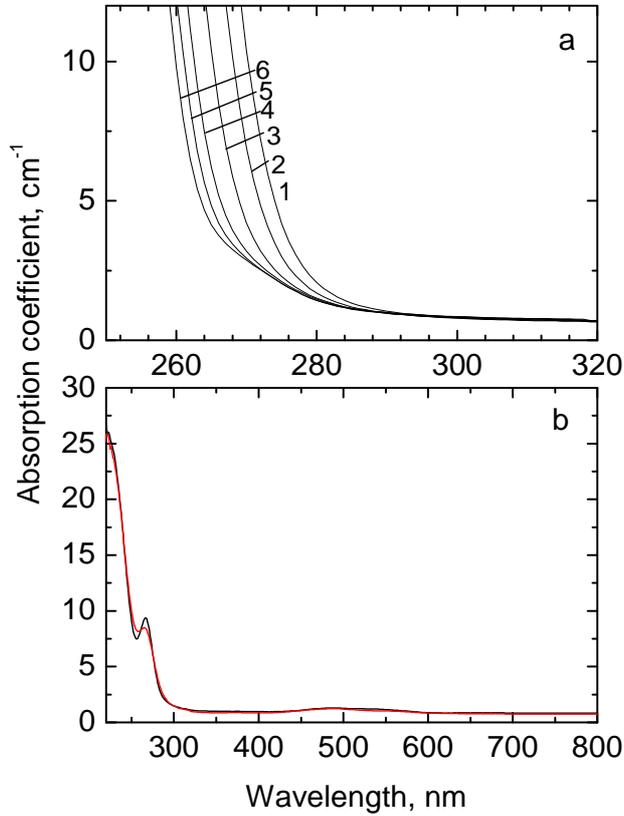

Fig. 4 (a) Temperature variation of absorption spectra of CaWO$_4$: 1- 295 K, 2 - 252 K, 3 - 203 K, 4 - 146 K 5 - 100 K, 6 - 8 K. The scale of the x-axis is zoomed for visualisation of the changes. (b) Absorption spectra of Al$_2$O$_3$-Ti (700 ppm) measured at T=295 (red) and 8 K (black).

The molybdate crystals evidence the effect of energy structure even further. There is a marked difference in the position of the absorption edge of CdMoO$_4$ and CaMoO$_4$ although both crystals adopt sheelite structure. This can be readily understood using the results of theoretical calculation of the energy structure of the crystals [66]. In CaMoO$_4$ 4d-states of Mo dominate the bottom of the conduction band whereas in CdMoO$_4$ there is noticeable contribution from Cd 5s to the conduction states that resembles the electronic structure inherent for CdWO$_4$. This explains the large difference in the energy gap of these crystals. From the position of the absorption edge of $ZnMoO_4$ it is also clear that a simple correlation between the cation radius and band gap energy does not hold in molybdates; their absorption properties should be considered in the framework of energy structure, requiring theoretical calculations.

Finally, we make a remark related to the cryogenic aspect of operation of scintillators. As was shown by numerous studies, the absorption coefficient of dielectrics obey the classical Urbach rule [67] suggesting a blue shift of the fundamental absorption edge with cooling. This is due to the broadening of the band gap of the crystal with decreasing temperature. As example of such a shift, data for $CaWO_4$ are displayed in fig. 4a. Therefore, cooling generally reduces the absorption of scintillation light in the volume of the crystal, favouring cryogenic application of intrinsic scintillators. In the case of doped scintillators, such as $Al_2O_3$-Ti, impurities or defects are responsible for the structure of the absorption edge. As can be seen from fig. 4b the temperature has only a minor effect and this feature of doped materials must be taken into account.

## 2.4. Scintillation characteristics and their temperature dependences

Like in any other application, the use of a material as a cryogenic scintillator in the first place relies on an efficient process of conversion of ionizing radiation into light. The measurement of the temperature dependence of the scintillation light output is a direct way to assess the feasibility of the material for this application. Moreover, studies of temperature dependence of decay times and the light yield give an opportunity to gain insight into the features of the scintillation process in the material, and to identify possible routes for improvements.

### 2.4.1. Calcium tungstate (CaWO₄)

Historically, calcium tungstate was the first scintillation target that was actively used in a cryogenic experiment searching for dark matter [15, 68]. This naturally evoked a large interest in this material and prompted a number of studies of luminescence [49, 57], lattice dynamics [69] and phonon propagation [70] down to cryogenic temperatures. First we studied the temperature variation of scintillation properties of $CaWO_4$ as well as many other materials over the 350 down to 9 K range using MPC [38, 21]. Since the temperature dependant processes affecting the emission of solids (energy transport and thermal quenching) appear not to change significantly with temperature below ~10 K it is expected that the results obtained at this temperature can provide instructive information on the major scintillation characteristics at millikelvin temperatures. Although quite consistent with the fundamentals of solid state physics, this intuitive notion needed experimental proof that has been eventually delivered when scintillation properties of $CaWO_4$ where investigated down to 0.020 K [26].

Figure 5a shows the variation of scintillation light output of $CaWO_4$ over the 0.020 – 350 K temperature range. At high temperatures (> 250 K) the light output is dominated by thermal quenching: as the probability of non-radiative decays increases strongly with temperature, the emission intensity decreases. From 20 to 250 K one observes only a small reduction in the light output. The enhancement observed at ~20 K can be attributed to the contribution from the radiative recombination of carriers captured by shallow traps. In the region between 0.02 and 10 K the light output remains constant within the error limits. The latter is the most important practical finding of this study, as it provides an explicit demonstration of the constancy of the light yield in the operating temperature range of CPSDs.

The temperature dependence of the decay time constant of $CaWO_4$ is shown in figure 5b. It illustrates the features of the scintillation kinetics in tungstates and molybdates with the general formula $AMeO_4$ (A=Ca, Mg, Zn, Cd; and Me=Mo, W). These materials are considered to be slow scintillators at room temperatures. The scintillation decay curve is usually presented as a sum of two or three exponential components as following:

$$f(t) = y_0 + \sum_i A_i \exp(-t/\tau_i), \tag{1}$$

where $y_0$ is background, $A_i$ and $\tau_i$ are amplitude and decay time constants. It should be noted that the decay process is mainly controlled by the component with the long decay time whose relative weight (product $\tau_i A_i$) is dominant. The decay time constant is of the order of a few microseconds at room temperature and increases by a few orders of magnitude when the crystal is cooled. Such behaviour of the long decay constant is a characteristic of the decay kinetics of tungstates, molybdates as well as other scintillators.

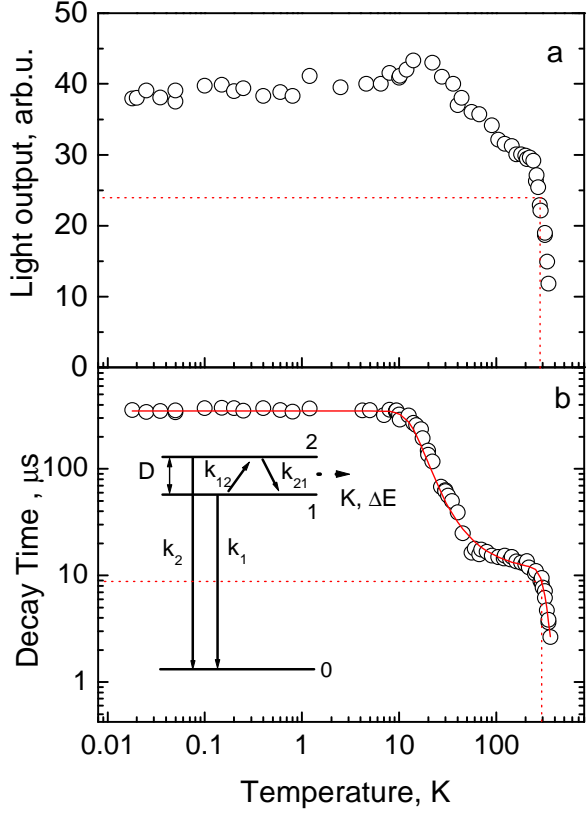

Fig. 5. (a) Temperature dependence of scintillation light output of CaWO$_4$ measured at excitation with 5.5 MeV α-particles from $^{241}$Am. (b) Temperature dependence of the long scintillation decay time constant of CaWO$_4$. The solid curve displays the result of the best fit to the experimental data using the three-level model shown in inset (Eq.2). Parameters of the fit are: $k_1 = 3.0 \times 10^3$ s$^{-1}$, $k_2 = 1.1 \times 10^5$ s$^{-1}$, $D = 4.4$ meV, $K = 8.6 \times 10^9$ s$^{-1}$, $\Delta E = 320$ meV. The dotted lines indicate the value of scintillation parameters at T=295 K.

The decay kinetic of CaWO$_4$ and its temperature dependence was analysed within the framework of a simple three-level model of the emission centre with one metastable level. The key features of this model have been suggested by Beard et al. [71] to provide a qualitative interpretation of the radiative decay in tungstates. Using the differential equations that describe the dynamics of population of the levels in the case of thermal equilibrium we derived the following expression for the temperature dependence of the decay time constant $\tau$ :

$$\frac{1}{\tau} = \frac{k_1 + k_2 \exp(-D/kT)}{1 + \exp(-D/kT)} + K \exp(-\Delta E / kT),\tag{2}$$

where $k_1$ and $k_2$ are the probabilities of radiative decay from levels *1* and *2* separated by an energy gap $D$, $K$ is the probability for the non-radiative decay rate and $\Delta E$ is the energy barrier of the non-radiative quenching process. It should be noted that the three-level model does not account for the energy transfer processes that provide additional channels for the radiative relaxation at moderate temperatures. Nonetheless, as can be seen from fig 5a, Eq. (2) provides adequate fitting to the experimental results. Therefore the model offers a sensible approximation, allowing to describe the features of the decay kinetics and to give the parameters of the relaxed excited state of the emission centre in CaWO$_4$.

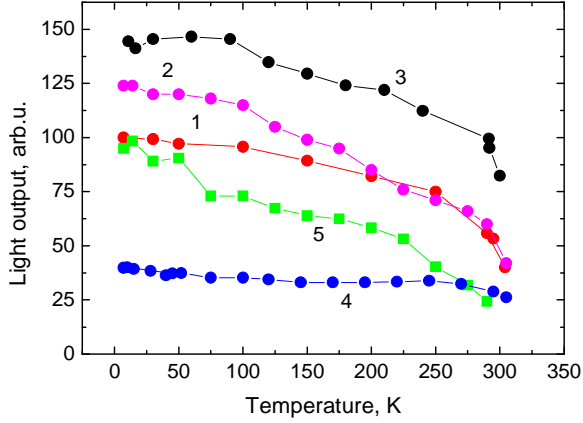

Fig.6. Temperature dependence of the scintillation light output in CaWO$_4$ (1), ZnWO$_4$ (2), CdWO$_4$ (3), MgWO$_4$ (4) and CaMoO$_4$ (5) measured at excitation with 5.5 MeV α-particles from [241]Am.

### 2.4.2. Other tungstates and molybdates

The results of these investigations of calcium tungstate proved that data from studies at ~10 K can be used to assess the suitability of scintillation materials for cryogenic applications. This provided the basis for benchmarking of all other materials. The variation of scintillation light yield of several tungstate and molybdate crystals in the temperature range 8-310 K is displayed in figure 6. The relative light yield was obtained from pulse height spectra by correcting for the spectral sensitivity of the photodetector. Table 1 summarises the results of our studies of the main scintillation properties of these as well as other potentially interesting materials. The measured dependences show a very similar trend in all crystals under study. In tungstates, at first, the light yield rapidly increases as the temperature is lowered to T~250-280 K. In terms of a classical model this roll-off temperature evidences the start of the thermally activated quenching process. Below this temperature the light yield exhibits moderate changes and at T<100 K it flattens. It should be noted that at low temperatures $ZnWO_4$ and $CdWO_4$ exhibit superior scintillation efficiency in comparison with calcium tungstate. The relative light output of optimised samples is about 120 and 145% with respect to $CaWO_4$ (see table 1). The molybdates exhibit qualitatively the same features but the temperature of the thermal quenching is considerably lower. $CaMoO_4$ is the only compound exhibiting a practically useful light yield at room temperature, about 55% that of $CaWO_4$ [43]. The scintillation light yield of cadmium molybdate is reasonably high only at low temperatures.

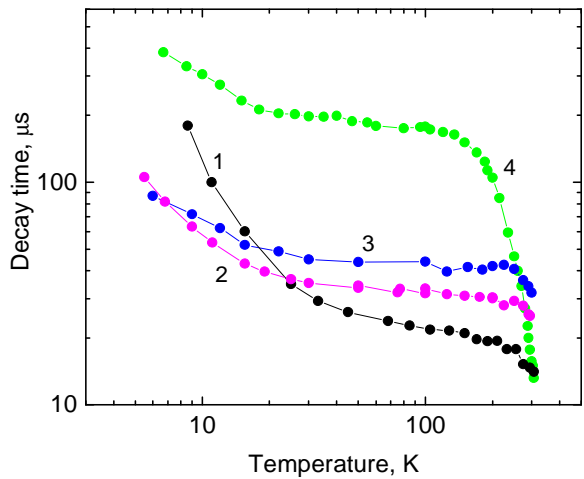

Fig. 7. Temperature dependence of the principal scintillation decay time constant of $CdWO_4$ (1), $ZnWO_4$ (2), $MgWO_4$ (3) and $CaMoO_4$ (4) measured at excitation with 5.5 MeV α-particles from [241]Am. The data are presented on a log-log scale for better visualisation of general trends with temperature.

Figure 7 shows the scintillation decay time constant of wolframite tungstates and $CaMoO_4$ measured over the 8-310 K temperature range. The long scintillation decay time constant changes by more than an order of magnitude with cooling to cryogenic temperatures.

Despite obvious differences, the dependencies exhibit several features which are common to all crystals. In the high temperature range the kinetics of the radiative decay is dominated by thermal quenching and the scintillation decay constant rapidly increases with cooling. As is seen from fig. 7 this part of the $\tau = f(T)$ dependence is especially noticeable in $CaMoO_4$. This is because the thermal quenching in calcium molybdate starts at much lower temperatures compared with tungstate crystals. Below the roll-off temperature the non-radiative processes are much less efficient and the increase of the decay time constant is less pronounced. The character of the scintillation kinetics changes abruptly below 20 K: it demonstrates considerable slowing of the decay. As is demonstrated for $CaWO_4$, this is due to the fine energy structure of the emitting centre in tungstates and molybdates that constitutes a metastable level a few meV below the emitting one. Since the experimental results do not extend below 5 K, we did not endeavour to map out the dynamics of decay processes by applying the model developed for $CaWO_4$ as it is strongly affected by the character of $\tau = f(T)$ dependence in the millikelvin range. It should be noted that relevant studies of the $CdWO_4$ and $ZnWO_4$ luminescence for optical excitation made [72] are consistent with our results and corroborate such model.

The scintillation decay time constants of the crystals studied at temperatures of 295 K and 9 K are listed in Table 1. An analysis of the temperature changes of the scintillation decay time constant shows that in the crystals with scheelite structure ($CaWO_4$, $CaMoO_4$ and $CdMoO_4$) it is substantially longer than in the tungstates with wolframite structure ($MgWO_4$, $ZnWO_4$ and $CdWO_4$). Apparently the magnitude of the effect that lifts the degeneracy of the emission levels and leads to the formation of a separate metastable state depends on the structure of the emission centre in the crystal. Overall the timing characteristics of the tungstates and molybdates currently pose no concern for applications in cryogenic rare event search. Nevertheless, the shorter decay time constants of wolframites can offer some advantage for pulse shape analysis – an option that is now being looked into.

Table 1. Scintillation properties of crystals for cryogenic rare event searches

| Crystal | Density, g/cm$^3$ | Emission peak, nm | Decay time, μs (α-particle excitation) | | Light output, (relative to CaWO$_4$ at 9K)* | |
|---|---|---|---|---|---|---|
| | | | 295 K | 9 K | 295 K | 9 K |
| $CaWO_4$ | 6.06 | 420 | 9 | 360 | 55 | 100 |
| $MgWO_4$ | 5.66 | 480 | 36 | 90 | 30 | 40 |
| $ZnWO_4$ | 7.87 | 490 | 24 | 110 | 60 | 120 |
| $CdWO_4$ | 7.90 | 480 | 14 | 180 | 80 | 145 |
| $CaMoO_4$ | 4.35 | 540 | 16 | 380 | 30 | 95 |
| $CdMoO_4$ | 6.07 | 550 | - | 460 | - | 80 |
| $Bi_4Ge_3O_{12}$ | 7.13 | 480 | 0.43 | 138 | 45 | 150 |
| $CaF_2$ | 3.18 | 280 | 1 | 930 | 15 | 60 |
| $MgF_2$ | 3.17 | 385 | - | 8900 | - | 70 |
| $Al_2O_3$-Ti | 3.98 | 290;430;740 | 0.15;...;3 | ... | 15 | 30 |
| ZnSe | 5.42 | 640 | - | 8** | - | 120 |

\* - light output in the units proportional to the number of emitted photons

\*\* - decay time constant represents the time interval leading to an intensity reduction by an order of magnitude ($\tau_{0.1}$).

Here we would like to remark on other tungstates and molybdates for which the scintillation properties were studied down to cryogenic temperatures, i.e. $PbWO_4$, $PbMoO_4$, $ZnMoO_4$ and $MgMoO_4$. The two last compounds, $ZnMoO_4$ and especially $MgMoO_4$ are very poor scintillators even at low temperatures. The low symmetry of the crystal structure and the high degree of disorder are believed to be the principal contributions responsible for the low luminescence efficiency of these compounds [73]. There is some scope for application of $ZnMoO_4$ in the search for neutrinoless 2β-decay [74] but $MgMoO_4$ is not anymore considered. Lead tungstate and molybdate have a low temperature of luminescence quenching and consequently exhibit very low luminescence yield at room temperature. This improves with cooling and at cryogenic temperatures the scintillation light output of pure $PbWO_4$ and $PbMoO_4$ is fairly high [28]. These crystals could certainly be used as cryogenic scintillators but at present excessive contamination by radioactive [210]Pb isotope spoils their application in rare event searches. Production of these materials with the required level of purity is certainly a path to pursue and such attempts are being undertaken by scintillator manufacturers.

### 2.4.3. Bismuth germanate (Bi₄Ge₃O₁₂)

$Bi_4Ge_3O_{12}$ is a classical example of an intrinsic scintillator that has been extensively examined at room temperatures but very little has been known about the low temperature scintillation properties of this compound. Recent studies of Moszynski et al [75] indicated a significant (factor of three) increase of the light yield at T=77 K. We measured the temperature dependence of the light output and the decay time of $Bi_4Ge_3O_{12}$ from room temperature down to 6 K [30]. The variation of the scintillation light output with temperature demonstrates the features that are common for intrinsic scintillators. As the temperature decreases to 100 K the scintillation light output of the crystal increases steadily (see fig. 8a). At lower temperature, however, the scintillation response remains virtually constant. The total light output of the scintillator increases by a factor of ~3.5 from T=295 K to 6 K. This estimate, based on the absolute room temperature value of scintillation efficiency [75], gives a light yield of $Bi_4Ge_3O_{12}$ at 6 K equal to 23700±2600 ph/MeV, which is that of 150% of $CaWO_4$ (see table 1).

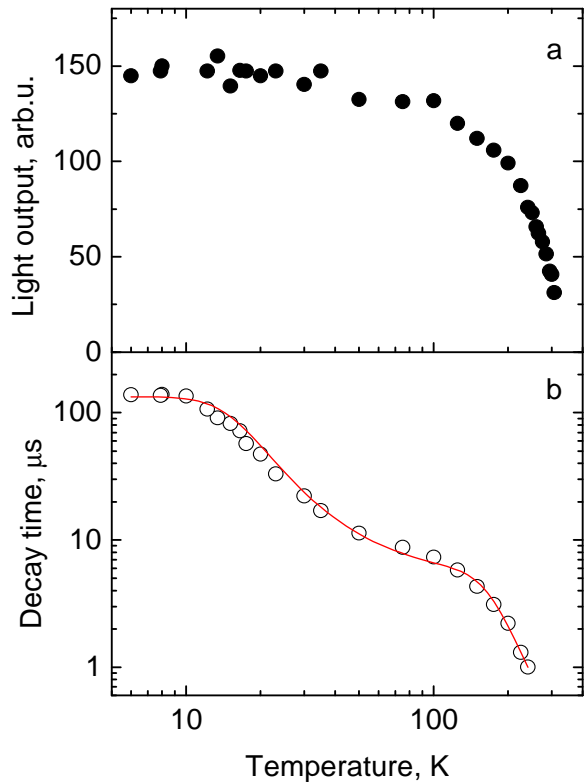

Fig. 8. (a) Temperature dependence of the scintillation light output of $Bi_4Ge_3O_{12}$. (b) Temperature dependence of the main scintillation decay time constant of $Bi_4Ge_3O_{12}$. The solid curve displays the result of the best fit to the experimental data using the three-level model (Eq.2). The parameters of the fit are: $k_1 = 7.5 \times 10^3 \text{ s}^{-1}$, $k_2 = 4.5 \pm 0.3 \times 10^5 \text{ s}^{-1}$, $D = 6.4 \pm 0.5$ meV, $K = 1.3 \pm 0.2 \times 10^8 \text{ s}^{-1}$, $\Delta E = 105 \pm 4$ meV.

Figure 8b shows the variation of the main decay time constants with temperature. The character of the temperature dependence of the decay time constant observed for $Bi_4Ge_3O_{12}$ is typical for the three-level model of the emitting centre constituting a ground and two excited levels of which the lower one is metastable. The luminescence of $Bi_4Ge_3O_{12}$ is attributed to the radiative transition within $Bi^{3+}$, and given the electron configuration of excited state of $Bi^{3+}$, it can be understood by referring to the model discussed above (Sec. 2.4.1) . The emission at low temperatures originates from the $^3P_0$ metastable level, while at higher temperature transitions occurs mainly between the excited $^3P_1$ and the ground $^1S_0$ states. The model predicts the existence of the low-temperature plateau in the $\tau = f(T)$ dependence and the gradual decrease of the decay time constant with temperature, caused by establishing a thermal equilibrium between the two emitting levels. The shortening of the decay time constant $\tau$ with increasing temperature $T$ can be understood as the result of a thermally-activated re-population of excited states and their non-radiative de-excitation.

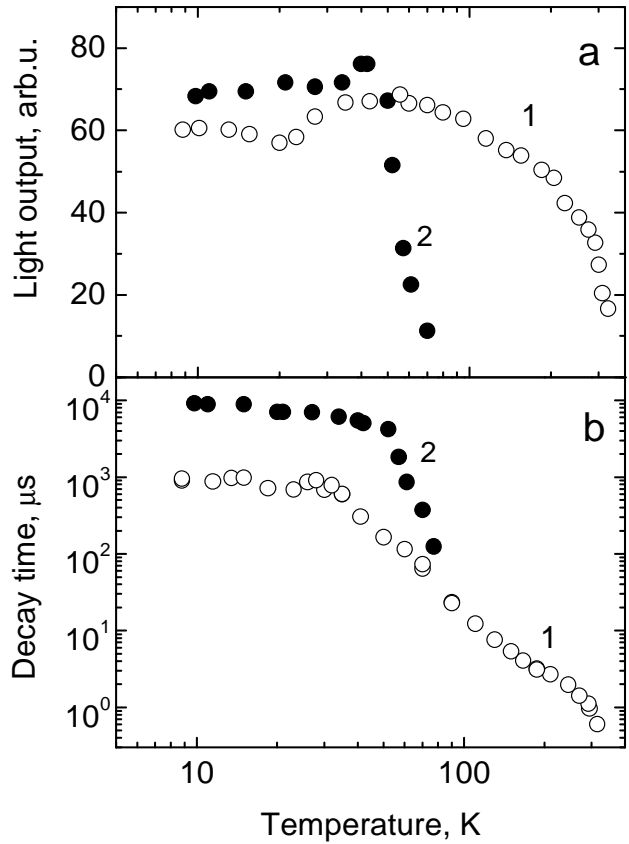

Fig. 9. (a) Temperature dependence of the scintillation light output in CaF$_2$ (1) and MgF$_2$ (2). (b) Temperature dependence of the main scintillation decay time constant of CaF$_2$ (1) and MgF$_2$ (2).

### 2.4.4. Fluorides

The results of comprehensive measurements of scintillation properties of pure CaF$_2$ and MgF$_2$ over a wide temperature range (9-300 K) are reported in [22] and [37]. The temperature variation of the scintillation light output and the main decay time constant in the crystals are shown in fig. 9a and b, respectively. It can be seen that calcium fluoride exhibits measurable scintillation at room temperature. The light output of the crystal increases about 2.5-fold with decreasing temperature. Magnesium fluoride, in contrast, shows no scintillation at room temperature. The emission appears after cooling the crystal to below 70 K, where it rises steeply, reaching the maximum at ca. 40 K. Because of the parity-forbidden character of the radiative transitions of the triplet STE, the fluorides exhibit very slow (~$10^{-3}$ s) decay time constant at low temperatures when it is not affected by thermal quenching. This is evidenced by the temperature dependence of the main decay time constants of $CaF_2$ and $MgF_2$, shown in fig. 9b. The long (microsecond) scintillation decay constants obtained in the present study are consistent with previous findings [60], suggesting that the scintillation of the fluoride crystals originates exclusively from the radiative transition of a triplet STE and that no impurities or defects are involved in the emission process. Estimates of the low-temperature light output of the fluoride relative to $CaWO_4$ give 60% and 70 % for $CaF_2$ and $MgF_2$, respectively (see table 1), indicating that  fluoride crystals are potentially good cryogenic scintillators. It is worthwhile recalling that $CaF_2$ is the material on which the concept of simultaneous detection of scintillation and phonon response for discrimination between the nuclear and electron recoils has been tested for the first time [12].

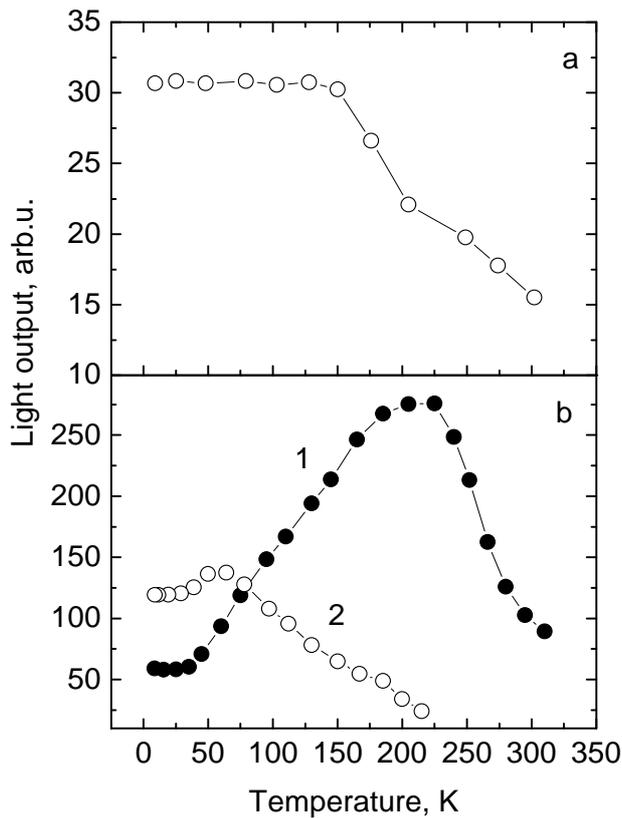

Fig.10. (a) Temperature dependence of the scintillation light output in $Al_2O_3$-Ti (700 ppm).
(b) Temperature dependence of the scintillation light output in ZnSe-Te (1) and pure ZnSe (2).

### *2.4.5 Ti-doped sapphire and ZnSe*

The studies of scintillation properties of ZnSe and $Al_2O_3$-Ti encountered a number of technical difficulties. The principal problem is that the main luminescence band of ZnSe and $Al_2O_3$-Ti are located in the red and near infrared spectral region (640 and 740 nm, respectively), where the sensitivity of conventional photomultipliers used for detection of scintillation is poor. Owing to this, the evaluation of the light yield is subjected to large errors. Furthermore these crystals exhibit complex decay kinetics. For ZnSe it is due to the recombination character of the luminescence whereas for Ti-doped sapphire it is due to a composite character of the emission: the bands peaking at 290, 430 and 740 nm are caused by different luminescence mechanisms that manifest essentially different timing characteristics; i.e. nanosecond decay for exciton emission, millisecond decay for $Ti^{3+}$ and microsecond decay for F-centre emission [63, 76]. In both cases a multi-exponential fit to such complex decay is problematic. Therefore, here we present merely the results on relative changes of the light output of these crystals with temperature. In Ti-doped sapphire the light yield increases gradually with cooling and remains constant below the roll-off temperature which is 150 K (see fig. 10a). The temperature changes are controlled by the processes of thermal quenching. The total increase of the light output of Ti-doped $Al_2O_3$ at cooling to 9 K is by a factor of two [31, 63]. The nominal concentration of titanium was found to be between 50 and 100 ppm; for this concentration the low-temperature light output of the $Al_2O_3$-Ti-doped sapphire is assessed to be 30% of $CaWO_4$.

To estimate the light output of zinc selenide we carried out comparative measurements of temperature dependences of the light yield for Te-doped and pure ZnSe. As can be seen from fig. 10b there is a marked dissimilarity in the dependences. This points to differences in the mechanisms that control the scintillation process in pure and doped crystals at different temperatures. In doped crystals, the scintillation efficiency is governed to a major extent by the efficiency of the energy transfer from the host to the emitting centres. Thus, the light output of ZnSe-Te first increases with cooling (stage of thermal quenching) then it decreases below T=225 K and, finally, below T=50 K it remains constant. The decrease of the light output of ZnSe-Te below the critical temperature of 225 K suggests that with cooling the probability of the energy transfer reduces, illustrating hereby a characteristic feature of doped scintillators. In contrast, pure ZnSe exhibits a temperature dependence of the light output which is typical for self-activated scintillators. According to fig. 10b the light output of ZnSe at low temperature is very close to what ZnSe-Te exhibits at room temperature. Based upon the room temperature value of the absolute scintillation yield of ZnSe-Te, 28300 ph/MeV [77] we estimated that the relative light yield of ZnSe at 9 K is in excess of CaWO$_4$ (see Table 1). The results of our low-temperature characterisation of zinc selenide evidenced that this material has very good prospects for cryogenic applications.

### 3. Discussion

### 3.1. Energy efficiency of scintillators in framework of phenomenological model

When the incoming high-energy photon or particle is absorbed in a material it creates many electrons and holes through the photoelectric effect, Compton scattering and pair production. Within very short time ($10^{-12}$ s) the electron-hole pairs become thermalised in the conduction and valence bands when the major part of their energy is transferred to phonons[1]. Only a small fraction of the deposited energy is transferred into scintillation. The conventional approach to the assessment of the efficiency of the scintillation phenomenon is based upon the semi-empirical theory of conversion of high energy radiation into light. The process is divided into three consecutive stages: absorption of the high-energy radiation and creation of charge carriers, energy transfer to the emission centre and finally, the luminescence stage [78, 79, 80, 81, 82]. Following this distinction the number of photons emitted by a scintillator can be expressed as the product of three parameters:

$$N_{ph} = N_{e-h} S Q, \tag{3}$$

where $N_{e-h}$ is the number of electron hole pairs created, $S$ is the probability of energy transfer to the luminescence centre and $Q$ is the luminescence quantum efficiency.

It is possible to derive $N_{e-h}$ from the energy balance of the process introducing the parameter of conversion efficiency $B$. It is defined as the ratio of electron-hole pairs created in the conversion process $N_{e-h}$ to their maximum possible number $N_{max}$:

$$B = \frac{N_{e-h}}{N_{max}}.$$

(4)

The value of $N_{max}$ can be calculated if one knows the minimum energy necessary to create a secondary electron-hole pair in the material with energy gap $E_g$. From the consideration based on the energy conservation law it is apparent that this energy should be $> 2E_g$ (minimum energy for electron-electron scattering or the Auger-process). Using a simple phenomenological model, Shockley [83] estimated this value for semiconductors to be $\sim 3E_g$. Later, based on the results of Monte Carlo simulations, Robins [84] derived that the lowest energy needed to produce a secondary electron-hole pair is equal to $2.35E_g$. This allows writing the expression for the number of electron-hole pairs produced by a high-energy quantum of energy $E_\gamma$:

$$N_{e-h} = N_{max}B = \frac{E_\gamma}{2.35E_g}B.$$

(5)

Substituting (5) into (3) gives the expression for the number of produced scintillation photons:

$$N_{ph} = N_{e-h}SQ = \frac{E_\gamma}{2.35E_g}BSQ.$$

(6)

Finally the equation for the energy efficiency of the scintillation process can be written as follows:

$$\eta = \frac{E_\lambda N_{ph}}{E_\gamma} = \frac{E_\lambda}{2.35E_g}BSQ,$$

(7)

where $E_\lambda$ is the energy of the scintillation photon[2].

---

[1] The energy carried by phonons can be subsequently measured via the "phonon channel" of CPSD. Though this channel is of primary importance for the operation of the CPSD as a radiation sensor its further discussion lies outside the scope of this work that deals exclusively with the scintillation aspect.

[2] In some recent publications this equation is written in a slightly different format:

$$\eta = \frac{E_\lambda}{\alpha E_g}SQ$$

with the comment that the parameter $\alpha$ is a phenomenological parameter varying between 2-3 [10], 2-4 [80] or 2-7 [5]. As such, this implies that $\alpha$ includes the conversion efficiency factor $B$ as a variable ($\alpha = 2.35/B$). In this work we follow the approach of [78] which allows gaining a better insight into the details of the energy conversion process in materials.

The expression for the conversion efficiency $B$ was derived by Lempicki et al [78, 79] on the basis of Robin's model of the energy conversion process in phosphors [84]. This model takes into account energy losses due to phonon-assisted thermalisation processes in different materials. It has been shown that the conversion efficiency can be expressed by the following relation [79]:

$$B = \frac{1}{1 + 0.65K} \qquad (8)$$

The parameter $K$, responsible for the losses, is defined as the ratio of energy lost by optical phonons to the energy lost for an ionisation event. This is a material dependent parameter that can be calculated as following [84]:

$$K = 0.244 \times 10^4 \left\{ \frac{1}{\varepsilon_\infty} - \frac{1}{\varepsilon_0} \right\} \frac{[\hbar\omega_{LO}]^{3/2}}{1.5E_g} . \qquad (9)$$

Here $\varepsilon_\infty$ and $\varepsilon_0$ are the high-frequency and static dielectric constants, $\hbar\omega_{LO}$ is the energy of the longitudinal optical phonons, dominating the rate of phonon-mediated energy loss in solids. Combining (8) and (9) with (7) gives the expression to start with for assessing the performance limits of different scintillation materials:

$$\eta = \frac{E_\lambda}{2.35E_g} \left[ 1 + 0.158 \times 10^4 \left\{ \frac{1}{\varepsilon_\infty} - \frac{1}{\varepsilon_0} \right\} \frac{[\hbar\omega_{LO}]^{3/2}}{1.5E_g} \right]^{-1} SQ \qquad (10)$$

The expression contains two unknown parameters, i.e. $S$ and $Q$, so such assessment is usually done under the assumption that the quantum efficiency $Q$ is equal to unity [78, 79, 80, 84]. This assumption is reasonably justified since for many materials of interest, specifically for activated phosphors, the quantum efficiency measured at room temperature is >0.9 [85, 86]. This seems to be the case, too, for activated scintillators: an absolute quantum efficiency of NaI-Tl at room temperature has been reported as 0.83±0.07 [87].

Concerning the probability of energy transfer $S$ it is commonly agreed that in general this is an unknown quantity, creating a large uncertainty when a comparison with experimental results is attempted. Though adequate models of the transfer stage have been developed for activated alkali halides [88] and Ce-doped scintillators [89, 90], theoretical analysis of this process provides no unambiguous, quantitative answer since the probability of capture for the exciton at the activator site is a function of excitation density and activator concentration. Nonetheless for the optimal concentration of activator in NaI-Tl and for low density of the initial excitations the value of $S$ is estimated as 0.89 [88]. Therefore it seems reasonable to assume that at least for some materials the transfer efficiency can be close to 1, which is adopted by some authors to estimate the maximum energy efficiency $\eta_{max}$ of scintillation materials [80, 84].

However, a more practical approach is to use experimental results to derive the transfer efficiency that permits to verify the correctness of the model and to assess the performance limit of the materials. The usefulness of this approach has been demonstrated in [78, 79] where the scintillation properties of a few materials where analysed and it was shown that the low transfer efficiency in several cases is the factor limiting the light output. It was also demonstrated that for few activated scintillators their performance is close to the limit defined by the condition that quantum efficiency and transfer efficiency approach unity. The data for three scintillators that illustrate this finding, i.e. NaI-Tl, CsI-Tl and LaCl$_3$-Ce, are listed in the Table 2. It should be noted that the physical parameters are taken from the Ref. [78] unless other indicated. We also used here the latest data on the measurements of scintillation light yield that should improve overall accuracy.

Taking into consideration that Tl-doped NaI and CsI have been around for a long time it is very likely that their performance reached the fundamental limit. Admittedly this is also true for LaCl$_3$-Ce, which has been subjected to extensive technological development over the last decade because of its commercial success. Altogether this gives us the encouraging indication that, despite of the semi-empirical character, the methodology used provides a useful framework for the treatment of the scintillation process in solids and yields reasonable results for activated scintillators.

Table 2. Parameters of scintillators.

| Material | T, K | $E_g$, eV | $\hbar\omega_{LO}$, meV | $\varepsilon_\infty$ [a)] | $\varepsilon_0$ | $B$ | $E_\lambda$, eV | $Q$ | $S$ | LY, ph/keV | $\eta$, % |
|----------|------|-----------|------------------------|---------------------------|-----------------|-----|-----------------|-----|-----|------------|-----------|
| NaI-Tl | 295 | 5.9 | 22 | 3.1 | 6.9 | 0.91 | 3.0 | 0.83[87] | 0.80 | 44[91] | 13.2 |
| CsI-Tl | 295 | 6.4 | 11 | 3.1 | 6.4 | 0.97 | 2.3 | 1[b)] | 0.88 | 57[91] | 13.1 |

| LaCl$_3$-Ce | 295 | 6.8[92] | 27[93] | 3.7[94] | 9.3[94] | 0.89 | 3.7 | 1 [b] | 0.83 | 48[91] | 17.8 |
|---|---|---|---|---|---|---|---|---|---|---|---|
| | | | | | | | | | | | |
| Bi$_4$Ge$_3$O$_{12}$ | 295 | 5.0 | 45 | 4.6 | 16.3 | 0.76 | 2.5 | 0.13[95] | 0.86 | 7.2[91] | 1.8 |
| | 80 | | | | | | | 0.44[95] | 0.79 | 22.7[75] | 5.7 |
| CaWO$_4$ | 295 | 5.2[57] | 108[96] | 3.5 | 10.4[96] | 0.41 | 2.9 | 0.76[97] | 0.61 | 15.8[98] | 4.6 |
| | 9 | | | | | | | 1 [b] | 0.85 | 28.7* | 8.3 |
| CaMoO$_4$ | 295 | 4.0[99] | 113[96] | 3.8 | 10.8[96] | 0.37 | 2.3 | 0.41[100] | 0.55 | 8.9[43] | 2.0 |
| | 9 | | | | | | | 1[b] | 0.87 | 27.3* | 6.2 |
| CdWO$_4$ | 295 | 4.2[65] | 112[101] | 5.0 | 17[102] | 0.43 | 2.6 | 0.76[b] | 0.83 | 27.4[103] | 7.1 |
| | 9 | | | | | | | 1 [b] | 0.90 | 39.6* | 10.2 |

*- our results; a) - calculated as $\varepsilon_\infty = n^2$; b) estimate.

### 3.2. Features of the scintillation process in intrinsic scintillators

The next step is the extension of this methodology to intrinsic (or self-activated) scintillators. In these materials luminescence is the property of the constituent elements of the host lattice which eliminates the need for the energy transfer. As a result of this, the parameter, characterising the probability of energy transfer ($S$), can be assumed to be unity, and that should reduce the uncertainty of results obtained from Eq. (8).

However there are two important issues to bear in mind while dealing with these materials. First is the temperature dependence of the quantum efficiency. This effect, which leads to non-radiative losses, can be quite substantial for many compounds and must be taken into account. The luminescence quantum efficiency is unity only at low temperatures and decreases with heating according to the classical Mott equation [104]:

$$Q = \frac{1}{1 + C \exp(-\Delta E / kT)}, \qquad (11)$$

where $C$ is the frequency constant that characterises the rate of the non-radiative decay, $\Delta E$ is the activation energy, $k$ is the Boltzmann constant and $T$ is the temperature. The bismuth germanate scintillator (Bi$_4$Ge$_3$O$_{12}$) is a classical example that can be used to demonstrate the importance of this effect. The light yield of this material at room temperature is fairly low and as can be concluded from inspecting Table 1 this is mainly due to a very low quantum efficiency (0.13). As the temperature decreases to 80 K the quantum efficiency increases and that explains the substantial increase of the light yield of $Bi_4Ge_3O_{12}$. The temperature variation of the quantum efficiency is an important issue which is directly related to the subject of this study.

Interestingly, the gain in light yield of $Bi_4Ge_3O_{12}$ is governed almost exclusively by the change of the quantum efficiency, indicating that the parameter $S$ varies only insignificantly with temperature. The results obtained in the framework of this model evidenced that the assumption of $S$ being close to unity is quite reasonable. However at this point we should recall an important second feature of intrinsic scintillators. The energy transfer process in these materials plays essentially a different role compared to that in activated scintillators. Since the luminescence centres are present in every lattice cell, the effect of energy transfer is a detrimental process as it allows the excitations to be captured by quenching centres and be dissipated without radiation. In a way this is an extreme example of concentration quenching in a lattice composed exclusively of luminescence ions. Thus, though mathematically equation (10) remains the same, the meaning of $S$ should be understood as a fraction of excitations reclaimed by luminescence centres. Such a feature of the energy transfer stage in intrinsic scintillators has a very important consequence: the lower probabilities of the energy transfer process in an intrinsic scintillator results in a higher probability for the excitations to be converted into scintillation light.

We can now consider the implications this issues impose upon the scintillation properties of intrinsic scintillators of our interest. First, it is useful to recall a few general notions from the energy transfer theory relevant to the interpretation of these results. It is generally acknowledged that exciton migration plays a decisive role in the energy transport process in both activated and intrinsic scintillators. The exciton migration is treated as a number of a single-step energy transfer processes in terms of a random walk model. The two most important single-step mechanisms of the energy transfer process are multipolar and exchange interactions. The rate of energy transfer between two ions is a function of distance $R$ between these ions [105]:

$$P_{dd} \propto (1/R)^6 \quad \text{(dipole-dipole interaction)} \tag{12}$$

$$P_{ex} \propto \exp(-R) \quad \text{(exchange interaction)} \tag{13}$$

The single-step mechanisms exhibit only weak temperature dependence, whereas the energy transfer controlled by excitons varies strongly with temperature; the temperature dependence is contained in the diffusion coefficient $D$ and is defined by a physical process that limits the mean free path of the exciton.

The details of the energy transfer process can be studied by measuring the temperature and concentration dependences of luminescence in activated materials. Neikirk and Powell [106] examined this process in $Bi_4Ge_3O_{12}$ doped with 1% $Er^{3+}$ and concluded that at low temperature the excitons are immobile and the transfer of energy to the activators is a single-step interaction process. At high temperature, thermally activated migration of excitons is responsible for the energy transfer. Nonetheless, the short diffusion length of excitons at room temperature (20 Å) makes the energy transfer process between activator ions rather inefficient. Note that in contemporary scintillation crystals the tolerable concentration of quenching centres is a few orders of magnitude lower than 1% [107]. Combining these facts we can put forward the conclusion that because of the short distance of exciton migration, the probability for excitations to reach the quenching centres is low at any temperature in $Bi_4Ge_3O_{12}$. This explains the high value of $S$ in this crystal.

### 3.3. Features of the scintillation process in tungstate and molybdate crystals: effect of structure.

Let us now consider the scintillation properties of tungstate and molybdate scintillators. Table 2 lists the parameters characterising the scintillation process in three crystals: $CaWO_4$, $CaMoO_4$, and $CdWO_4$. These crystals are selected because we were able to collect all parameters that are necessary for the assessment of scintillation performance. Inspection of Table 2 shows that the conversion efficiency $B$ is very similar in all tungstates and slightly lower in $CaMoO_4$. Apparently, the parameters $Q$ and $S$ are responsible for the variation in the scintillation light yield. Fortunately, for $CaWO_4$ and $CaMoO_4$ it is possible to put forward a few suggestions using the experimental results on measurements of the luminescence quantum efficiency. Vyshnevskyi et al [100] showed that the quantum efficiency of $CaMoO_4$ at room temperature is 0.41 and at cooling below 100 K it maintains a constant value. The quantum efficiency of $CaWO_4$ at room temperature was reported by Botden [97] to be 0.76 and inspection of the temperature change of the quantum efficiency allows to conclude that it is >0.9 below 200 K. Using these values and the measured light yield of the crystals we calculated the parameter $S$, characterising the efficiency of the energy transfer stage.

As can be seen from Table 2, the room temperature values of $S$ in $CaWO_4$ and $CaMoO_4$ are very similar but noticeably lower than unity (0.61 and 0.55 respectively). This is a good indication that a significant fraction of the excitations is channelled to the quenching centres. On the other hand, the low-temperature value of $S$ is much higher and close to that of $Bi_4Ge_3O_{12}$. These findings can be readily explained using the results of earlier studies of the energy transfer process in $CaWO_4$. Botden [97] investigated the luminescence of Sm-doped tungstates and noticed that in $CaWO_4$ the efficiency of the energy transfer from the host lattice to impurity ions reaches 0.94 at high temperature. According to the results of Treadaway and Powell [108] the exciton diffusion length in $CaWO_4$ is 120 Å at room temperature. Using this number one can roughly estimate that for the scenario of isotropic random walk, excitons should experience capture during their motion if the concentration of trapping centres is about 8 ppm. We recall that the concentration of impurities in a $CaWO_4$ crystal, for which the scintillation light yield was measured, is a few tens of ppm [98], which should be reasonable cause for the energy losses. As the temperature decreases, the exciton migration ceases and the energy is transferred only to the nearest neighbour trough the single-step interaction process [108], being consistent with the observed enhancement of $S$.

The model of energy transfer through exciton migration can explain the observed features in the scintillation characteristics of intrinsic scintillators. This can be further illustrated using $CdWO_4$ as example. Since there are no published data on the luminescence quantum efficiency of cadmium tungstate based on the similarity of the temperature dependence of the light output with $CaWO_4$ (see fig. 6 and also [97]), we made the assumption that this parameter should be similar in both crystals. Calculated in such a way the value of $S$ at high and low temperatures is very high (0.83 and 0.90), indicating that the probability of non-radiative losses of energy in this crystal is especially low. Clearly, the energy transfer to trapping centres in $CdWO_4$ is inhibited. This is consistent with the studies of [97] the luminescence of Sm-doped tungstates evidencing that the efficiency of activator emission in $CdWO_4$ and $ZnWO_4$ at any temperature is significantly lower compared with that of $CaWO_4$-Sm [97]. Therefore it is now reasonable to state that in terms of efficiency of energy transfer there is a substantial difference between crystals with scheelite ($CaWO_4$ and $CaMoO_4$) and wolframite structure ($CdWO_4$ and ZnWO$_4$). This difference is believed to be related to the crystal structure of the materials and specifically to how the luminescence centres - oxyanion [MeO$_n$] complexes - are linked.

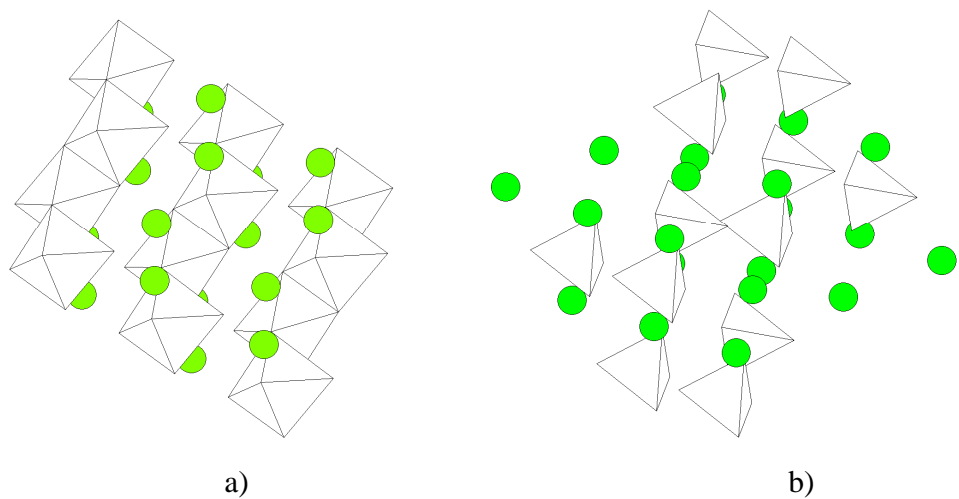

a)                      b)

Fig. 11. Fragments of the crystal structure of wolframite (a) and scheelite (b). Octahedrons in (a) and tethraedrons in (b) represent WO$_6$ and WO$_4$ oxyanion complexes, respectively. The positions of metal cations are shown as green spheres.

The crystal structure of wolframite tungstates contains linear zigzag chains of tungstate octahedrons connected through the edge sharing (see fig. 11a). The distance between the neighbouring W$^{6+}$ ions in the chain is noticeably shorter than it is between the chains (3.30 and 5.02 Å respectively for CdWO$_4$). Therefore the wolframite structure exhibits a strong anisotropy of the physical properties. Regarding the luminescence process, we have a system with a significant difference in the energy transfer rate along the chain of the oxyanion WO$_6$ complex when compared with the perpendicular direction. The assessments made for CdWO$_4$, using equations (12) and (13), indicate that the energy transfer rate along the chain is 5.5 times higher in the case of exchange interaction. The difference is even more impressive (by a factor of 30) if one considers a dipole-dipole mechanism of interaction. This means that in tungstates with wolframite structure the energy transfer predominantly occurs in one direction, which in turn limits the chances of excitation to reach the quenching centre. In the crystals with scheelite structure (CaWO$_4$ and CaMoO$_4$) W$^{6+}$ ions occupying a central position in the primitive cell are coordinated tetrahedrally by four oxygen ions. The distance between the neighbouring tungsten ions is the same in all directions (3.87 Å) and hence the structure can be considered as a three-dimensional. The energy transfer process can take place in any direction with the same probability that inevitably fosters a trapping process. Qualitatively this is well supported by the observations. Data of table 2 shows that, in comparison with cadmium tungstate, $CaWO_4$ and $CaMoO_4$ manifest lower values of $S$, which is evidence of enhanced energy losses during the transfer stage. The difference in the parameter $S$ for two types of crystal structure is more pronounced at high temperature due to activation of the exciton migration process that enhances the energy transfer. That is why, as far as scintillation properties are concerned, the wolframite crystal structure is much more favourable in comparison with the scheelite structure.

### 4. Concluding remarks

Operation of scintillation detectors at cryogenic temperatures is a new area of scientific and technological research that is attracting growing interest. The quest for efficient cryogenic scintillators motivated the investigation of many traditional ($CaWO_4$, $CdWO_4$, $ZnWO_4$, $Bi_4Ge_3O_{12}$) and less-conventional scintillators ($CaMoO_4$, $Al_2O_3$-Ti, $CaF_2$) as well as materials that have not been considered as scintillators hitherto ($ZnSe$, $MgF_2$). We gave an up-to-date overview on the low-temperature scintillation properties of these compounds. Regarding the major parameter of interest for cryogenic application, the scintillation efficiency, tungstates and molybdates can be ranked by the scintillation light output at low temperature (highest one first): $CdWO_4$–$ZnWO_4$–$CaWO_4$–$CaMoO_4$–$CdMoO_4$–$MgWO_4$. Cadmium tungstate is definitely the most efficient scintillator over wide temperature range and it is proven to be an excellent choice for cryogenic experiment searching for neutrinoless double-beta decay [20, 74]. However, the intrinsic radioactivity of the β-active [113]Cd spoils its merit of application in dark matter searches. Obviously the same caveat applies to cadmium molybdate.

Calcium tungstate is a scintillator currently used in cryogenic dark matter search. The use of this scintillator facilitated the development of the CPSD technique for a dark matter search and it is expected that $CaWO_4$ as well as $CaMoO_4$ should be the constituent parts of the multi-target absorber in a future experiments [19]. However it is now realised that due to constrains imposed by the chemistry of calcium, the purification of Ca-based compounds to the level of intrinsic radioactivity acceptable for future experiments is a significant challenge. Therefore it is now believed that $ZnWO_4$ holds a biggest promise for dark matter search [109]. This is because zinc tungstate, which has been produced as a commercial scintillator for a while, has reached maturity and the material composition favours low intrinsic radioactivity. The encouraging results of recent developments in the technology of fabrication of glued phonon sensors [33, 110], as well as impressive progress in the production of large zinc tungstate crystals of high quality [34] underpin the good prospects for application of this material in future cryogenic dark matter experiments.

We have employed semi-empirical theory of conversion of high-energy radiation in light of analysing the efficiency of the scintillation process. We have highlighted that in self-activated scintillators the process of energy transfer has a more detrimental effect upon the scintillation efficiency as opposed to doped scintillators. This singular significance of the transfer step permits to explain some important features of self-activated scintillators, such as temperature behaviour or the effect of the crystal structure. It has been shown that the value of parameter $S$ that characterises the efficiency of the energy transfer stage in self-activated scintillators is close to unity in some compounds at low temperatures, indicating that their scintillation efficiency approaches the theoretical limit. We have analysed how the structure of tungstate crystals affects the scintillation properties and deduced that the features of the crystal structure of wolframites impose a limitation upon the energy transport transforming it into essentially a one-dimensional structure. This is a principal reason for better performance characteristics of $CdWO_4$ and $ZnWO_4$ (wolframite structure) in comparison with $CaWO_4$ and $CaMoO_4$ (scheelite structure) where the energy transfer process is fairly isotropic. Altogether this allows us to make a practical conclusion: the same efforts aiming at improving the scintillation properties of these crystals will lead to different results. The performance of $CdWO_4$ and $ZnWO_4$ wolframites may be noticeably improved for example through the reduction of defects and quenching impurities while optimisation of scintillating $CaWO_4$ and $CaMoO_4$ is a more difficult and subsequently costly task.

Low-temperature scintillation studies were carried out for a selected group of other promising compounds which hitherto have not been investigated. Due to their high scintillation yield at low temperatures, $Bi_4Ge_3O_{12}$ and ZnSe are very appealing materials for cryogenic applications. The CPSDs with respective targets have already been realised and valuable scientific results, such as discovery of α-decay of [209]Bi [17] were delivered. Furthermore, $Bi_4Ge_3O_{12}$ might become an important target permitting a cross comparison with detectors that use a germanium semiconductor as target [111]. However due to contamination by natural Bi radiative isotopes this material will need extensive purification to reach the required level of radiopurity before it can be used in future experiments searching for rare events. On the other hand, recent studies of CPSD with ZnSe confirmed the great promise of this material for future cryogenic experiments searching for neutrinoless 2β-decay of $^{82}$Se [74].

Fluorides and sapphire represent target materials that are particularly favourable to search with for the spin-dependent interaction of WIMPs due to 100% abundance of nuclei with non-zero spin. Measurements of scintillation properties of undoped $CaF_2$ and $MgF_2$ over a wide temperature range evidenced that both compounds exhibit high scintillation yield at low temperature. Excellent phonon propagation properties of $Al_2O_3$ underpinned the use of this material as cryogenic detector in the first generation of cryogenic experiments searching for dark matter [112]. The extensive studies of Ti-doped $Al_2O_3$ carried out recently demonstrated that it is an adequate low-temperature scintillator and it is now being tested as CPSD in an experimental setup for dark matter search [29, 113]. The feasibility to use fluorides and $Al_2O_3$-Ti as CPSDs opens a possibility to explore new scenarios of interaction in dark matter search.

Summing up, in this study we have shown that there are quite a few scintillation materials that demonstrate good performance at low temperatures and as such they are well suited for applications as CPSD. Currently this application encompasses merely the specialised field of search for rare event where particle identification and high energy resolution are indispensable. However it is also realised that the high sensitivity and discrimination power of CPSD offers a great promise beyond. Whatever future application is thought, it will require significant improvement of the performance of identified scintillators as well as new materials for specific purposes.

## Acknowledgments

This work was supported by grants from the Science and Technology Facilities Council (STFC) and the Royal Society (London).

## References

[1] T. A. Edison, Nature, 53 (1896) 470.
[2] W. Crookes, Proc. Roy. Soc. London, 71 (1903) 405–408.

[3] E. Rutherford, Phil. Mag. 21 (1911) 669-688.

[4] R. Hofstadter, Phys. Rev., 75 (1949) 796.

[5] M.Weber, J. Lumin., 100 (2002) 35-45.

[6] C. W. E. van Eijk, Phys. Med. Biol., 47 (2002) R85-R106.

[7] P. Lecoq, A. Annenkov, A. Gektin et al., *Inorganic Scintillators for Detector Systems: Physical Principles and Crystal Engineering*, Springer-Verlag, Berlin, Heidelberg, 2006.

[8] P. A. Rodnyi, *Physical Processes in Inorganic Scintillators*, CRC Press, Boca Raton, 1997.

[9] M. D. Birosumoto and P. Dorenbos, Phys. Stat. Sol A, 206 (2009) 9-20.

[10] M. Nikl, Meas. Sci. Technol., 17 (2006) R37-R54.

[11] *Status and Perspectives of Astroparicle Physics in Europe* http://www.aspera-eu.org/images/stories/files/Roadmap.pdf.

[12] A. Alessandrello, V. Bashkirov, C. Brofferio et al., Phys. Let. B, 420 (1998) 109-113.

[13] L. Gonzalez-Mestres and D. Perret-Gallix, Nucl. Instr. Meth. Phys. Res. A, 279 (1989) 382-387.

[14] C. Bobin, I. Berkes, J. P. Hadjout et al., Nucl. Instr. Meth. Phys. Res. A, 386 (1997) 453-457.

[15] P. Meunier, M. Bravin, B. Bruckmayer et al., Appl. Phys. Let., 75 (1999) 1335-1337.

[16] G. Angloher, M. Bauer, I. Bavykina et al., Astropart. Phys, 31 (2009) 270-276.

[17] P. de Marcillac, N. Coron, G. Dambier et al. , Nature, 422 (2003) 876-878.

[18] C. Cozzini, G. Angloher, C.Bucci et al., Phys. Rev. C, 70, (2004) 064606.

[19] H. Kraus, M.Bauer, I. Bavykina et al., Nucl Phys. B- Proc. Suppl., 173 (2007) 168-171.

[20] L. Gironi, C. Arnaboldi, S. Capelli et al., Opt. Mater., 2008 in press.

[21] V. B. Mikhailik and H. Kraus, J. Phys: D: Appl. Phys., 39 (2006) 1181-1191.

[22] V. B. Mikhailik, H. Kraus, J. Imber and D. Wahl, Nucl. Instr. Meth. Phys. Res. A, 566 (2006) 522-525.

[23] S. Pirro, J. M. Beeman, S. Capelli et al., Phys. At. Nucl., 69 (2006) 2109-2116.

[24] H. Kraus, V. B. Mikhailik and D. Wahl, Radiat. Meas., 42 (2007) 921-924.

[25] F. A. Danevich, S. Henry, H. Kraus et al., Phys. Stat. Sol A, 205 (2008) 335-339.

[26] V. B. Mikhailik, H. Kraus, S. Henry and A. J. B. Tolhurst, Phys. Rev. B, 75 (2007) 184308.

[27] A. N. Annenkov, O. A. Buzanov, F.A. Danevich et al., Nucl. Instr. Meth. Phys. Res. A, 584 (2008) 334-345.

[28] L. L. Nagornaya, F.A. Danevich, A.M. Dubovik et al., IEEE Trans. Nucl. Sci., 56 (2009), 2513-2518.

[29] P. C. F. Di Stefano, N. Coron, P. de Marcillac et al., J. Low Temp. Phys., 151 (2008) 902-907.

[30] J. Gironnet, V. B. Mikhailik, H. Kraus et al., Nucl. Instr. Meth. Phys. Res. A, 594 (2008) 358-361.

[31] M. Luca, N. Coron, C. Dujardin et al., Nucl. Instr. Meth. Phys. Res. A, 606 (2009) 545-551.

[32] H. Kraus, V. B. Mikhailik, Y. Ramachers et al., Phys. Lett. B, 610 (2005) 37–44.

[33] I. Bavykina, G. Angloher, D. Hauff et al., Opt. Mater., 31 (2009) 1382-1387.

[34] L. L. Nagornaya, B.V. Grinyov, A. M. Dubovik et al., IEEE Trans. Nucl. Sci. 56 (2009) 994-997.

[35] F. A. Danevich, D. M. Chernyak, A. M. Dubovik, et al., Nucl. Instr. Meth. Phys. Res. A, 608 (2009) 107-115.

[36] P. C. F. Di Stefano and F. Petricca, Opt. Mater., 31 (2009) 1381.

[37] Proceedings of the 1$^{st}$ Int. Workshop "Radiopure Scintillators for EURECA" (RPScint'2008) arXiv:0903.1539v2 [nucl-ex].

[38] H. Kraus, V. B. Mikhailik and D. Wahl, Nucl. Instr. Meth. Phys. Res. A, 553 (2005) 522-534.

[39] V. B. Mikhailik and H. Kraus in Proceedings of the 1st Int. Workshop "Radiopure Scintillators for EURECA" (RPScint'2008) arXiv:0903.1539v2 [nucl-ex] pp.64-71.

[40] R.H.Gilette, Rev. Sci. Instrum., 21 (1950) 294-301.

[41] C. E. Tyner and H.G. Drickamer, J. Chem. Phys. 67 (1977) 4103-4115.

[42] B. C. Grabmaier, IEEE Trans. Nucl. Sci., NS-**31** (1984) 372-376.

[43] V. B. Mikhailik, S. Henry, H. Kraus and I. Solskii, Nucl. Instr. Meth. Phys. Res. A, 583 (2007) 350-355.

[44] F. A. Kröger, *Some Aspects of the Luminescence of Solids*, Elsevier, New York, 1948

[45] G. Blasse Structure and Bonding 42 (1980) 1-41.

[46] E. G. Reut, Izv. Akad. Nauk SSSR Ser.Fizicheskaya, 49 (1985) 2032-2038 (in Russian).

[47] V. N. Vishnevsky, V.N. Kulitsky, M.S. Pidzyrailo et al., Ukrainskyi Fizychnyi Zhurnal, 17 (1972) 1225-1230 (in Ukrainian).

[48] V. B. Mikhailik, H. Kraus, M. Itoh et al., J. Phys. Cond. Matt., 17 (2005) 7209-7218.

[49] V. B. Mikhailik, H. Kraus, G. Miller et al., J. Appl. Phys., 97 (2005) 083523.

[50] Y. Zhang, N. A. W. Holthwarth and R. T. Williams, Phys. Rev. B, 57 (1998) 12738.

[51] M. Itoh, N. Fujita and Y. Inabe, J. Phys. Soc. Jap., 75 (2006) 084705

[52] M. Nikl, V.V. Laguta and A. Vedda, Phys. Stat. Sol. B, (2008) 1701-1722.

[53] M. Bacci, S. Porciani, E. Mihokova et al., Phys. Rev. B., 64 (2001) 104302.

[54] Yu. A. Hizhnyi, T. N. Nikolaenko and S. G. Nedilko, Phys. Stat. Sol. C, 4 (2007) 1217-1221.

[55] M. Itoh and T. Sakurai, Phys. Rev B 73 (2006) 235106.

[56] B. Henderson and G. F. Imbush, *Optical Spectroscopy of Inorganic Solids*, Clarendon Press, Oxford, 1989

[57] V.B. Mikhailik, H. Kraus, D. Wahl et al., Phys. Rev. B, 69 (2004) 205110.

[58] J. H. Beaumont, W. Hayes, D. L. Kink, G. P. Summers, Proc. Royal. Soc. London A, 315 (1970) 69-97.

[59] T. I. Nikitinskaya, P. A. Rodnyi and S. B. Mikhrin, Opt. Spectrosc. 39 (1975) 411-413.

[60] R. T. Williams, C. L. Morquart, J. W. Williams and M. N. Kabler, Phys. Rev. B, 15 (1977) 5003.

[61] N. G. Zakharov, T.I.Nikitinskaya and P.A.Rodnyi, Sov. Phys. Sol. State, 24 (1982) 709-710.

[62] V. Ryzhikov, N. Stargynskiy, V. Seminozhenko et al, Functional Materials, 11 (2004) 61-66.

[63] V. B. Mikhailik, H. Kraus, M. Balcerzyk et al., Nucl. Instr. Meth. Phys. Res. A, 546**,** (2005) 523-534.

[64] D. Wahl, V. B. Mikhailik and H. Kraus, Nucl. Instr. Meth. Phys. Res. A, 570 (2007) 529-535.

[65] R. Lacomba-Perales, J. Ruiz-Fuertes, D. Errandonea, Europhys. Lett., 83 (2008) 37002

[66] M. Fujita, M. Itoh, T. Katagiri et al., Phys. Rev. B. 77 (2008) 155118

[67] F. Urbach, Phys. Rev., 92 (1953) 1324.

[68] M. Bravin, M.Bruckmayer, C.Bucci et. al., Astropart. Phys., 12 (1999) 107-114.

[69] A. Senyshyn, H. Kraus, V. B. Mikhailik and V. Yakovyna, Phys. Rev. B., 70 (2004) 214306

[70] K. Hayasaka, R. Higashi, J. Suda, et al., Sol. State Commun., 143 (2007) 386-389.

[71] G. B. Beard, W. H. Kelly and M. L. Mallory, J. Appl. Phys., 33 (1962) 144-147.

[72] V. Babin, P. Bohacek, E. Bender, et al., Radiat. Measur., 38 (2004) 533-537.

[73] V. B. Mikhailik, H. Kraus, V. Kapustyanyk et al., J. Phys. Cond. Matt., 20 (2008) 365219.

[74] S. Pirro, 6th ILLIAS Annual Meeting, Dresden 16 - 19 February 2009, http://ilias.in2p3.fr/ilias_site/meetings/documents/ILIAS_6th_Annual_Meeting/180209_Pirro.pdf

[75] M. Moszynski, M. Balcerzyk, W. Czarnacki et al., IEEE Trans. Nucl. Sci., 51 (2004) 1074-1079.

[76] A. I. Surdo, V. A. Pustovarov, V. S. Kortov et al., Nucl. Instr. Meth. Phys. Res. A, 543 (2005) 234-239.

[77] M. Balcerzyk, W. Klamra, M. Moszynski et al., Nucl. Instr. Meth. Phys. Res. A, 482 (2002) 720-727.

[78] A. Lempicki, A. J. Wojtowicz and E. Berman, Nucl. Instr. Meth A, 333 (1993) 304-311

[79] A. Lempicki and A. J. Wojtowicz, J.Lumin., 60/61 (1994) 942-947.

[80] G. Blasse, J. Lumin., 60/61 (1994) 930-935.

[81] P. A. Rodnyi, P. Dorenboss, C. W. E. van Eijk Phys. Stat Sol. B, 187 (1995) 15-29.

[82] A. N. Vasil'ev, Procc of the 8th Int. Conf. on Inorganic Scintillators and their use in Scientific and Industrial applications, Alushta, Ukraine, 2005, pp. 1–6.

[83] W. Shockley, Solid State Electronics, 2 (1961) 35-60.

[84] D. J. Robins, J.Electrochem. Soc., 127 (1980) 2694-2702.

[85] *Phosphor Handbook*, ed by S. Shionoya and W. M.Yen, CRC Press, Boca Raton, 1999.

[86] *Luminescence: From Theory to Applications*, ed by C. Ronda, Wiley-VCH, 2007.

[87] V. N. Vyshnevskyi and M. S. Pidzyrailo, Ukrainskyi Fizychnyi Zhurnal, 12 (1967) 1466-1473 (in Ukrainian).

[88] R. B. Murray and A. Meyer, Phys. Rev., 122 (1961) 815-826.

[89] A. J. Wojtowicz, M. Balcerzyk, E. Berman et al., Phys. Rev. B., 49 (1994) 14880-14895

[90] G. Bizarri and P. Dorenbos, Phys. Rev B., 75 (2007) 184302.

[91] J. T. M. de Haas and P.Dorenboss, IEEE Trans. Nucl. Sci. 55 (2008) 1086-1092

[92] J. Andriessen, O. T. Antonyak, P. Dorenbos et al., Opt. Commun. 178 (2000) 355–363.

[93] C. K. Asawa, R. A. Satten, and O. M. Stafsudd, Phys. Rev., 168 (1968) 957.

[94] D. W. Berreman and F. C. Unterwald, Phys. Rev., 174 (1968) 791.

[95] V.Y. Ivanov, A.V. Kruzhalov, V.A. Pustovarov et al., Nucl. Instr Meth A, 261 (1987) 150-152.

[96] A. S. Barker, Phys. Rev. B, 135 (1964) A742-A747.

[97] Th. P. J. Botden, Philips. Res. Rep., 6 (1951) 425-473.

[98] M. Moszyński, M. Balcerzyk, W. Czarnacki et al., Nucl. Instr. Meth. Phys. Res. A, 553 (2005) 578-591.

[99] M. Fujita, M. Itoh, S. Takagi et al., Phys. Sta. Sol. b 243 (2006) 1898-1907.

[100] V.N. Vyshnevsky, V.N. Kulytsky, M.S. Pidzyrailo, et al., Ukrainskyi Fizychnyi Zhurnal, 17 (1972) 1225-1230 (in Ukrainian).

[101] Z. Burshtein, S. Morgan, D. O. Henderson, et al., J. Phys. Chem. Sol., 49 (1988) 1295.

[102] V. N. Shevchuk and I. V. Kayun, Rad. Meas., 42 (2007) 847.

[103] M. Moszynski, M. Balcerzyk, M; Kapusta et al., IEEE Trans. Nucl. Sci., 52 (2005) 3124-3128.

[104] N. F. Mott, Proc. Royal. Soc. London A, 167 (1938) 384-391.

[105] D. L. Dexter, J. Chem. Phys., 21 (1953) 836-850.

[106] D. P. Neikirk and R. C. Powell, J. Lumin., 20 (1979) 261-270.

[107] B. Grynyov, V. Ryzhikov, J. K. Kim, M. Jae, *Scintillator crystals, radiation detectors and insruments on their base*, Ukraine, Kharkiv, 2006

[108] M. J. Treadaway and R. C. Powell, Phys. Rev. B., 11 (1975) 862-874.

[109] H. Kraus, F.A. Danevich, S. Henry et. al, Nucl. Instr. Meth. Phys. Res. A, 600 (2009) 594-598

[110] M. Kiefer, G. Angloher, M. Bauer, et al., Opt. Mater., 31 (2009) 1410-1414.

[111] A. Benoit, L. Bergé, J. Blümer, Phys. Lett. B, 616 ( 2005) 25-30.

[112] G. Angloher, M. Bruckmayer, C. Bucci, et al., Astropart. Phys., 18, (2002) 43-55.

[113] A. Calleja, N. Coron, E. García et al., J. Low Temp. Phys., 151 (2008) 848–853.